\documentclass{PoS}
\usepackage{bbm}
\usepackage{amssymb,amsmath}

\newcommand{\mnu}{\mathcal{M}_\nu}
\newcommand{\diag}{\mbox{diag}\,}

\title{Neutrino Physics - Models for Neutrino Masses and Lepton Mixing}

\ShortTitle{Neutrino Physics}

\author{\speaker{Grimus Walter} \\
        University of Vienna, Austria\\
        E-mail: \email{walter.grimus@univie.ac.at}}

\abstract{In these lecture notes we present mechanisms for the
          generation of Majorana neutrino masses and lepton mixing. 
          We consider simple extensions of the Standard Model. 
          Apart from a section about radiative mass generation, we put
          special emphasis on the seesaw mechanism and $\mu$--$\tau$
          interchange symmetry. 
          }

\FullConference{School on Particle Physics, Gravity and Cosmology\\
                 August 21 -September 2,  2006 \\
                 Dubrovnik}

\begin{document}

\section{Introduction}

\paragraph{Scope of the lecture notes:}
These notes comprise two hours of lecture. The style is rather
elementary and detailed. Therefore, only a few subjects have been
presented and the choice of subjects is rather subjective, reflecting
the preferences of the author.
The notes are intended to offer access to some sections of model building for
neutrino masses and lepton mixing from where the interested reader can
go on with more advanced literature.

General introductions to neutrino physics, with emphasis on neutrino
oscillations, are found in~\cite{reviews}. For recent developments in
models for neutrino masses and mixing see~\cite{theory-reviews}.
Ref.~\cite{lecture} contains aspects of both, phenomenology and theory,
and could be used as reading complementary to the present notes.
In Refs.~\cite{reviews,theory-reviews,lecture} one can find extended
bibliographies, whereas here we confine ourselves to work closely
related the presented subjects.
For reviews of the results of neutrino oscillation experiments
see~\cite{exp-reviews}. 

\paragraph{Introductory remarks:}
The results of the neutrino oscillation experiments have shown that at least
two neutrinos are massive and 
lepton mixing exists in analogy to quark mixing. 
What does this important finding---one of the most
spectacular discoveries in the recent history of particle 
physics---mean for
model building? To obtain a reasonable perspective one should 
take into account the following remarks. 
It is no problem to accommodate neutrino masses and mixing, 
the problem is rather to explain its characteristic features. 
Note that in the quark sector the mass and mixing problem is still
unsolved, after so many decennia and despite numerous attempts. 
It could very well be that the mass problem is decoupled from mixing
problem, i.e., perhaps one can find models which explain mixing but not the
masses. The mass problem could be more fundamental than the mixing problem in
the following sense. Some mixing angles might find an explanation in mass
ratios, see for instance the conjecture that the Cabibbo angle is
approximately given by $\sin \theta_c \simeq \sqrt{m_d/m_s}$ where $m_d$ and
$m_s$ are the masses of the down and strange quark,
respectively~\cite{wilczek}. Some of the
mixing angles in the lepton sector might be, in a first approximation,
independent of fermion masses, 
for instance the atmospheric mixing angle $\theta_\mathrm{atm}$ might be 
$45^\circ$ and thus maximal, and the small angle $\theta_{13}$ could zero. 
The meaning of these angles will
be explained in Section~\ref{majorana}.

As for accommodating neutrino masses and mixing, one could add to the
multiplets of the Standard Model right-handed neutrino singlets $\nu_R$, just
as one has right-handed quark singlets in the SM, and require conservation of
the total lepton number $L$. Then one would have in the lepton sector a
complete analogy to the quark sector, with massive Dirac neutrinos and lepton
mixing. 

Why are we not happy with this picture? 
Each of the three series of charged fermions (up quarks, down quarks and
charged leptons) has a strong hierarchy in the masses. For instance, in the
up quark sector we have $m_t/m_u \sim 4 \times 10^4$. Moreover, the
quark (CKM) mixing matrix is not extremely far from the unit matrix. 
This is in accord with the idea that both the quark mass hierarchy and the CKM
matrix being close to unity are founded in hierarchical structures 
of the up and down quark mass matrices.\footnote{This could of course be
  merely a prejudice.} On the other hand, in the lepton sector it is true
there is a strong hierarchy in the charged lepton masses with 
$m_\tau/m_e \sim 3500$, however, when compared with the neutrino masses,
experiments tell us that the largest neutrino mass is about six orders
magnitude smaller than the electron mass. Thus the relation between charged
lepton masses and neutrino masses is very different from the relation between
down and up quark masses. Furthermore, it came as a surprise that the lepton
mixing or PMNS matrix is very far from unity. Consequently, 
one would like to understand the following issues:
\begin{enumerate}
\item
Why are neutrino masses much smaller than charged lepton masses? \\
There are two generic ``proposals'' for a solution of this problem:
\begin{itemize}
\item The seesaw mechanism, 
\item radiative neutrino masses.
\end{itemize}
\item
Can one reproduce the special features of neutrino masses and lepton mixing?
The special features are
\begin{itemize}
\item[F1: ] 
a solar mixing angle of $\theta_\odot \simeq 34^\circ \!\!
\raisebox{0.7mm}{$
\begin{array}{c} \scriptstyle +3^\circ \\[-3mm] \scriptstyle -2^\circ 
\end{array}$},$ which is large but non-maximal, 
\item[F2: ] 
an atmospheric mixing angle of
$\theta_\mathrm{atm} \simeq 45^\circ \pm 8^\circ$,
which is large and perhaps maximal,
\item[F3: ] 
a small element $U_{e3}$ of the lepton mixing matrix, with
$|U_{e3}|^2 \equiv s_{13}^2 \lesssim 0.02$, 
\item[F4: ] 
a small ratio of neutrino mass-squared differences 
$\Delta m_\odot^2/\Delta m_\mathrm{atm}^2 \sim 0.03$.
\end{itemize}
\end{enumerate}
The approximate ranges for the mixing angles refer to 90\% confidence level
and are taken from~\cite{exp-reviews}.

\paragraph{The framework:} 
The framework of the lectures is defined by the following assumptions:
\begin{itemize}
\item
We consider simple extensions of the lepton sector of the SM, i.e.,
the considered gauge group is $G = SU(2)_L \times U(1)_Y$.
\item
As for extensions of the fermion sector, we take into account the addition
of right-handed neutrino singlets. 
\item
We discuss all possible extensions of the scalar sector, compatible
with $G$.
\item
We use flavour symmetries for enforcing certain features of the PMNS
matrix. 
\item
We assume Majorana nature of the neutrinos.
\end{itemize}
At this point we emphasize that an important question is 
whether the explanation of the features F1--4 is independent of the
general fermion mass problem. Here we assume that this is the
case. Otherwise, one would necessarily have to start with Grand Unified
Theories, where quark masses and the CKM matrix is inseparably
connected with the problem of lepton masses and the PMNS matrix.
Another interesting question, to be solved by future experiments, is 
how close to $45^\circ$ is the atmospheric mixing angle and how close
to zero is the small angle $\theta_{13}$. For the time being, sizeable
deviations of these angles from these values are allowed.
However, if it turns out that the atmospheric mixing angle is very
close to maximal and $\theta_{13}$ very close to zero, this 
could hint at a non-abelian flavour symmetry.

The plan of the lecture notes is as follows. 
In Section~\ref{majorana} we discuss Majorana mass terms and the
parameterization of lepton mixing. Section~\ref{extensions} introduces
extensions of the SM by right-handed neutrino singlets, together with
the seesaw mechanism, and extensions by additional scalar multiplets.
In Section~\ref{mu-tau-section} we consider the so-called
$\mu$--$\tau$-symmetry in the neutrino mass matrix. A model which
realizes this symmetry is constructed in Section~\ref{model}. Finally,
in Section~\ref{multi} 
we present some general considerations about about multi-Higgs models,
with soft breaking of lepton numbers in the mass terms of the
right-handed neutrino singlets.

\section{Majorana neutrinos and lepton mixing}
\label{majorana}

\paragraph{Majorana mass terms:}
We begin with some algebra. 
All spinors used here are 4-spinors.
In the space of 4-spinors the
charge-conjugation matrix $C$ is defined by 
\begin{equation}\label{C}
C^{-1} \gamma_\mu C = -\gamma_\mu^T,
\end{equation}
where the $\gamma_\mu$ denote the Dirac matrices. We will always work
in a representation of the Dirac matrices where $\gamma_0$ is hermitian
and the $\gamma_j$ ($j=1,2,3$) are anti-hermitian. The properties of
$C$ are
\begin{equation}\label{Cprop}
C^T = -C, \quad C^\dagger = C^{-1}.
\end{equation}
Whereas the first property follows from Eq.~(\ref{C}) alone, the
second one takes into account the hermiticity assumption.

The charge-conjugation operation is defined by
\begin{equation}\label{CC}
\psi^c \equiv C {\bar \psi}^T = C \gamma_0^T \psi^*.
\end{equation}
The projectors 
\begin{equation}
P_L \equiv (\mathbbm{1} - \gamma_5)/2, \quad
P_R \equiv (\mathbbm{1} + \gamma_5)/2
\end{equation}
produce so-called chiral 4-spinors. A spinor $\psi_L$ is called
left-handed if $P_L \psi_L = \psi_L$. Then, using 
$C^{-1} \gamma_5 C = \gamma_5^T$, it is easy to show that
\begin{equation}
P_R \left( \psi_L \right)^c = \left( \psi_L \right)^c.
\end{equation}
Thus, a charge-conjugate chiral spinor has the opposite chirality. 

A mass term is a Lorentz-invariant\footnote{For fermions it is
  actually invariance under $SL(2, \mathbbm{C})$, the covering group of
  the proper orthochronous Lorentz group.} bilinear 
in the Lagangian. A Dirac mass term has the structure 
$\bar\phi_R \psi_L + \mbox{H.c.}$, with independent chiral spinors
$\phi_R$, $\psi_L$. Note that in the mass term different chiralities
are necessary, otherwise the mass term would be identically zero. If
we have only one chiral spinor $\psi_L$ at our disposal, we can use
Eq.~(\ref{CC}) to form a right-handed spinor. In this way, we obtain a 
Majorana mass term (see also~\cite{lecture})
\begin{equation}\label{majterm}
- \frac{1}{2} m 
\left( \overline{\left( \psi_L \right)^c} \psi_L + \mbox{H.c.} \right) =
\frac{1}{2} m \psi_L^T C^{-1} \psi_L + \mbox{H.c.}
\end{equation}
The spinors transform as 
\begin{equation}
\psi_L \to \exp \left( 
-i \alpha_{\mu\nu} \sigma^{\mu\nu}/4 \right) \psi_L
\quad \mbox{with} \quad
\sigma^{\mu\nu} = \frac{i}{2}\,\left[ \gamma^\mu, \gamma^\nu \right]
\end{equation}
under Lorentz transformations, where the transformations are 
parameterized by the six coefficients of the real antisymmetric matrix 
$\left( \alpha_{\mu\nu} \right)$. 
Invariance of the term~(\ref{majterm})
is guaranteed because 
\begin{equation}
\exp \left( -i \alpha_{\mu\nu} \sigma^{\mu\nu}/4 \right)^T C^{-1} 
\exp \left( -i \alpha_{\lambda\rho} \sigma^{\lambda\rho}/4 \right) = C^{-1}
\quad \mbox{due to} \quad
C \left( \sigma^{\mu\nu} \right)^T C^{-1} = -\sigma^{\mu\nu}.
\end{equation}
The factor $1/2$ in Eq.~(\ref{majterm}) is necessary in order to
interpret $m$ as the mass appearing in the Dirac equation because this
factor is canceled in the functional derivative of the Lagrangian
with respect to
$\psi_L$ since it occurs twice (see second part of Eq.~(\ref{majterm})).

\paragraph{Mass term for Majorana neutrinos:}
According to the discussion above, we write down a Majorana neutrino
mass term
\begin{equation}\label{mnu}
\mathcal{L}_\mathrm{maj} = 
\frac{1}{2}\, \nu_L^T C^{-1} \mnu \nu_L + \mbox{H.c.},
\end{equation}
where $\nu_L$ contains an arbitrary number of left-handed neutrino fields.
Since fermion fields are anticommuting and $C$ is antisymmetric, we
have 
$\nu_{aL}^T C^{-1} \nu_{bL} = \nu_{bL}^T C^{-1} \nu_{aL}$,
where $\nu_{aL}$, $\nu_{bL}$ are 4-spinor neutrino fields occurring in
$\nu_L$. Therefore, the neutrino mass matrix $\mnu$ is a
\emph{symmetric} matrix, i.e.
\begin{equation}\label{symmetric}
\mnu^T = \mnu,
\end{equation}
which is complex in general.

The diagonalization of this mass matrix proceeds according to a
theorem, first proven by Schur~\cite{schur}: For any symmetric and
complex matrix $\mnu$ there exists a unitary matrix $V$ such that 
\begin{equation}\label{V}
V^T \mnu V = \hat m \equiv \mbox{diag}\,( m_1, m_2, m_3 ),
\end{equation}
where the $m_j$, the neutrino masses, are real and non-negative.
The matrix $V$ can be decomposed as
\begin{equation}\label{V1}
V = e^{i\hat \varphi}\, U\, \mbox{diag}\,(e^{i\rho}, e^{i\sigma}, 1),
\end{equation}
with a diagonal phase matrix $e^{i\hat \varphi}$.
The neutrino mass eigenfields $\hat\nu_L$ are then given by the relation
$\nu_L = V \hat\nu_L$, with Majorana fields and mass term given by
\begin{equation}\label{eigen}
\nu_M = \hat\nu_L + \left( \hat\nu_L \right)^c, \quad
\mathcal{L}_\mathrm{maj} = 
\frac{1}{2}\, {\hat\nu}_L^T C^{-1} \hat m \hat\nu_L + \mbox{H.c.},
\end{equation}
respectively.
The Majorana fields $\nu_M$ fulfill 
$\left( \hat\nu_M \right)^c = \hat\nu_M$.

Let us from now on assume that we work in a 
basis where the charged lepton mass matrix is diagonal. Then the phases in 
$e^{i\hat \varphi}$ are unphysical in lepton mixing because they 
can be absorbed into the left-handed charged lepton fields in the
following way. Consider the charged-current Lagrangian
\begin{equation}
-\mathcal{L}_{cc} = 
\frac{g}{\sqrt{2}}\, W^-_\mu \bar\ell_L \gamma^\mu \nu_L +
\mbox{H.c.} = 
\frac{g}{\sqrt{2}}\, W^-_\mu \bar\ell_L \gamma^\mu V \hat\nu_L +
\mbox{H.c.}
\end{equation}
The charged-lepton fields $\ell = \ell_L + \ell_R$ are Dirac fields,
thus their mass term is $\bar\ell_R \hat m_\ell \ell_L + \mbox{H.c.}$,
where $\hat m_\ell$ is the diagonal mass matrix. Then, defining new
fields $\ell'_L = e^{-i\hat\varphi} \ell_L$ and $\ell'_R =
e^{-i\hat\varphi} \ell_R$, $e^{i\hat\varphi}$ disappears from
$\mathcal{L}_{cc}$ without making reappearance in the mass term.

Note that the phase factors $e^{i\rho}$, $e^{i\sigma}$ of $V$ are
\emph{physical} for Majorana neutrinos and the phases 
$\rho$, $\sigma$ are called Majorana phases. We cannot absorb them in
the neutrino fields, just as we absorbed $e^{i\hat\varphi}$ in the
charged-lepton fields. 
If we absorb them in $\hat\nu_{1L}$, $\hat\nu_{2L}$, we shift
the masses in Eq.~(\ref{eigen}) according to 
$m_1 \to e^{-2i\rho} m_1$, $m_2 \to e^{-2i\sigma} m_2$ and the new
fields are not in the mass eigenbasis.

\paragraph{The lepton mixing matrix:}
According to the discussion before, the lepton mixing matrix is given by
\begin{equation}
U_\mathrm{PMNS} = U\,\diag (e^{i\rho}, e^{i\sigma}, 1).
\end{equation}
Using the convention of~\cite{RPP}, we decompose the unitary matrix
$U$ as 
\begin{equation}\label{U}
U = U_{23} U_{13} U_{12} =
\left( \begin{array}{ccc} 1 & 0 & 0 \\
0 & c_{23} & s_{23} \\ 
0 & -s_{23} & c_{23} \end{array} \right)
\left( \begin{array}{ccc} c_{13} & 0 & s_{13} e^{-i\delta} \\
0 & 1 & 0 \\ -s_{13} e^{i\delta} & 0 & c_{13} \end{array} \right)
\left( \begin{array}{ccc} c_{12} & s_{12} & 0 \\
-s_{12} & c_{12} & 0 \\ 0 & 0 & 1 \end{array} \right).
\end{equation}
We use the abbreviations $c_{23} \equiv \cos \theta_{23}$, etc.
The angle $\theta_{23}$ is also called atmospheric mixing angle
because it is the protagonist in atmospheric and long-baseline
$\nu_\mu \leftrightarrow \nu_\tau$ oscillations, with corresponding
mass-squared difference $\Delta m^2_\mathrm{atm}$. The mixing angle
$\theta_{13}$, for which only an upper bound exists, is responsible for
$\nu_e \leftrightarrow \nu_\mu$ oscillations. The angle $\theta_{12}$
appears in solar or very long-baseline oscillations; the latter are at
present only realized in the KamLAND experiment~\cite{exp-reviews}.
The corresponding mass-squared difference can always be chosen as 
$\Delta m^2_\odot = m_2^2 - m_1^2$ with $m_2 > m_1$.
The phase $\delta$ is analogous to the CKM phase and can, in principle, 
be probed in neutrino oscillations~\cite{reviews}. 

With the convention $m_2 > m_1$, there are two physically distinct
cases for $m_3$: $m_1 < m_2 < m_3$, the ``normal'' spectrum, and 
$m_3 < m_1 < m_2$, the ``inverted'' spectrum. In both cases, 
$\Delta m^2_\mathrm{atm}$ can be chosen as the largest mass-squared
difference. 

\section{Extensions of the SM} 
\label{extensions}

\subsection{Right-handed neutrino singlets} 
\label{right-handed}

\paragraph{Multiplets:}
In this section we extend the set of SM fields by right-handed neutrino
singlets. Thus we have $G$-multiplets with the following quantum numbers:
$$
\begin{array}{cccll}
& SU(2) & \!\!\times\!\! & \! U(1) & \\
D_L & \underline{\frac{1}{2}} && Y = -1 & \mbox{left-handed doublets,} \\
\ell_R & \underline{0} && Y = -2 & \mbox{right-handed charged lepton singlets,}
\\ 
\nu_R  & \underline{0} && Y = \hphantom{-}0 & 
\mbox{right-handed neutrino singlets,} \\
\phi   & \underline{\frac{1}{2}} && Y = \hphantom{-}1 & \mbox{Higgs
  doublet,} \\
\tilde\phi  & \underline{\frac{1}{2}} && Y = -1 & \mbox{Higgs doublet.} 
\end{array}
$$
The irreducible representations of $SU(2)$ are denoted by weak
isospin, $Y$ is the hypercharge.
The scalar doublet $\tilde\phi$ is not an independent degree of freedom. 
It is related to $\phi$ by
\begin{equation}
\tilde \phi \equiv i\tau_2 \phi^* 
\quad \mbox{or} \quad 
\phi = \left( \begin{array}{c} \phi^+ \\ \phi^0 \end{array} \right) 
\Leftrightarrow \tilde
\phi = \left( \begin{array}{c} {\phi^0}^* \\ -\phi^- \end{array} \right), 
\end{equation}
where $\tau_2$ is the second Pauli matrix and in the second part of the
equation we have assumed---without loss of generality---that the lower
component in $\phi$ is the one with zero electric charge. Clearly, the
hypercharge of $\tilde \phi$ is opposite to the hypercharge of $\phi$, but
under $SU(2)$ both fields transform in the same way: suppose $U \in SU(2)$ and
$\phi \to U \phi$, then also $\tilde\phi \to U \tilde\phi$. The reason is that
\begin{equation}\label{su2prop}
U^\dagger i\tau_2 U^* = i\tau_2 \Leftrightarrow 
U^T i\tau_2 U = i\tau_2,
\end{equation}
which is a special $SU(2)$ property.

There are two motivations for introducing the $\nu_R$. First of all,
in the SM the left-handed quark doublet fields have right-handed
$SU(2)$-singlet partners. From that point of view there is no reason
for omitting the $\nu_R$. Before the discovery of neutrino 
masses, the omission of $\nu_R$ was fine because in that way the
neutrinos stayed massless. The second motivation comes from GUTs
based on the gauge group $SO(10)$. In such models all chiral fermions
of one family are contained in the 16-dimensional irreducible spinor
representation.\footnote{This is actually a representation of its
  covering group $Spin(10)$.} Let us do the counting of 
the chiral fields per family:
\begin{center}
$2 \times 2 \times 3$ (quarks: up, down; L, R; colour) + $2 \times 2$ 
(leptons: $\ell$, $\nu$; L, R) = 16 
\end{center}
Thus, in $SO(10)$ GUTs the $\nu_R$ is automatically included. For an
introduction into $SO(10)$ GUTs see for instance~\cite{pal}.

\paragraph{The Lagrangian:} Let us assume that we have the multiplets
of the SM plus one $\nu_R$ per family; in addition, we allow 
for violations of all lepton numbers, including the total lepton
number $L$, and an arbitrary number of Higgs doublets. Then the Lagrangian
is given by 
\begin{equation}\label{Lseesaw}
\mathcal{L} = \cdots
- \sum_j \left[ 
\bar \ell_R \phi_j^\dagger \Gamma_j + 
\bar \nu_R {\tilde\phi}^\dagger \Delta_j \right] D_L
+ \mbox{H.c.} 
+ \left( \frac{1}{2}\, \nu_R^T C^{-1} \! M_R^* \nu_R + \mbox{H.c.} \right),
\end{equation}
where the dots indicate the gauge part. The $\nu_R$ mass term is of
Majorana form and is present because we allow for $L$-violation. The
requirement of $L$ conservation would forbid that term and lead to
Dirac neutrinos. In analogy to Eq.~(\ref{symmetric}), we have 
$M_R = M_R^T$. Spontaneous symmetry breaking of the SM gauge group
induces the mass matrices 
\begin{equation}\label{M}
M_\ell = \sum_j v_j^\ast \Gamma_j \,,
\quad 
M_D    = \sum_j v_j \Delta_j
\quad \mbox{with} \quad 
\langle \phi_j^0 \rangle_0 = v_j,
\end{equation}
where $M_\ell$ is the mass matrix of the charged leptons and the $v_j$
are the vacuum expectation values (VEVs) of the Higgs doublets. The matrix
$M_D$ goes together with $M_R$ to form a Majorana mass term for
left-handed neutrino fields~\cite{seesaw1}:
\begin{equation}\label{numass}
\mathcal{L}_{\nu\,\mathrm{mass}} = 
\frac{1}{2}\, \omega_L^T C^{-1} \mathcal{M}_{D+M}\, \omega_L + \mbox{H.c.},
\end{equation}
with the $6 \times 6$ matrix 
\begin{equation}\label{D+M}
\mathcal{M}_{D+M} = \left( \begin{array}{cc}
0 & M_D^T \\ M_D & M_R     \end{array} \right)
\quad \mbox{and} \quad \omega_L = 
\left( \begin{array}{c} \nu_L \\ C (\bar\nu_R)^T \end{array} \right).
\end{equation}

Let us sketch the derivation of Eqs.~(\ref{numass}) and (\ref{D+M}). 
We have to reformulate all mass terms with $(\nu_R)^c$. For this
purpose we reformulate Eq.~(\ref{CC}) as
\begin{equation}\label{(1)}
\nu_R^* = -C^{-1} \gamma_0 (\nu_R)^c,
\end{equation}
from $(\nu_R)^{cc} = \nu_R$ we derive 
\begin{equation}\label{(2)}
\bar \nu_R = \overline{(\nu_R)^{cc}} = -[(\nu_R)^c]^T C^{-1}
\end{equation}
and from Eq.~(\ref{C}) we obtain
\begin{equation}\label{(3)}
\gamma_0^T C^{-1} \gamma_0 = -C^{-1}.
\end{equation}
Using Eq.~(\ref{(2)}) we treat first the Dirac term:
\begin{equation}\label{Di}
-\bar\nu_R M_D \nu_L = 
[(\nu_R)^c]^T C^{-1} M_D \nu_L = 
\frac{1}{2} \left\{ [(\nu_R)^c]^T C^{-1} M_D \nu_L + 
\nu_L^T C^{-1} M_D^T [(\nu_R)^c] \right\}.
\end{equation}
Dealing with the Majorana term, we first take its complex conjugate,
then use Eq.~(\ref{(1)}) and finally Eq.~(\ref{(3)}):
\begin{eqnarray}
&& \left( \frac{1}{2}\, \nu_R^T C^{-1} \! M_R^* \nu_R \right)^\dagger = 
\frac{1}{2}\, \nu_R^\dagger C M_R \nu_R^* 
= \frac{1}{2}\, [-C^{-1} \gamma^0 (\nu_R)^c]^T 
C M_R [-C^{-1} \gamma^0 (\nu_R)^c] = \nonumber \\ && \label{Ma}
\frac{1}{2}\, [(\nu_R)^c]^T \left(- {\gamma^0}^T C^{-1} \gamma^0 \right) 
M_R (\nu_R)^c = 
\frac{1}{2}\, [(\nu_R)^c]^T C^{-1} M_R (\nu_R)^c.
\end{eqnarray}
Equations~(\ref{Di}) and (\ref{Ma}) are in exactly the form we
to have.

\paragraph{The seesaw mechanism:}
In the mass matrix~(\ref{D+M}), $M_D$ is generated by the VEVs of the Higgs
doublets, therefore, its elements are at most of the order of the electroweak
scale. On the other hand, the scale of $M_R$ is not protected 
by the gauge symmetry and there is no reason why it cannot be much larger. 
Indeed the basic assumption of the seesaw mechanism~\cite{seesaw} is 
$m_D \ll m_R$, where $m_D$ and $m_R$ are the scales of $M_D$ and
$M_R$, respectively.
A more precise formulation of this assumption is that the
largest eigenvalue of $\sqrt{M_D^\dagger M_D}$ is much smaller than the
smallest eigenvalue of $\sqrt{M_R^\dagger M_R}$.

To derive the seesaw mechanism, we are looking for a unitary $6 \times 6$
matrix $W$ which disentangles small from large scale. 
In the derivation we follow Ref.~\cite{disent} and make the ansatz
\begin{equation}\label{W}
W = \left( \begin{array}{cc} \sqrt{\mathbbm{1} - B B^\dagger} & B \\
-B^\dagger & \sqrt{\mathbbm{1} - B^\dagger B} 
\end{array} \right),
\end{equation}
such that
\begin{equation}\label{disent}
W^T \mathcal{M}_{D+M} W =
\left( \begin{array}{cc} \mnu & 0 \\ 0 & \mnu^\mathrm{heavy} 
  \end{array} \right)
\end{equation}
The right-hand side of this equation expresses the disentanglement of
small and large scales.
The matrix $W$ is modeled after a $2 \times 2$ rotation matrix.
The square root is understood as Taylor expansion  
$\sqrt{1-x} = 1 - \frac{1}{2}\,x - \frac{1}{8}\, x^2- \cdots$. 
Before we go on we do some parameter counting. The matrix $B$ is a general
complex $3 \times 3$ matrix and thus has 18 real parameters, and $W$
has the same number of parameters. A general unitary $6 \times 6$
matrix has 36 parameters. Thus $W$ has 18 parameters less and it is impossible
to diagonalize $\mathcal{M}_{D+M}$ with $W$. The lack of 18 parameters agrees
with the form of the matrix on the right-hand side of
Eq.~(\ref{disent}): $\mnu$ and $\mnu^\mathrm{heavy}$ are both
symmetric and 
would need each a unitary $3 \times 3$ matrix for diagonalization,
thus $2 \times 9 = 18$ further parameters.

Equation~(\ref{disent}) determines 
$B$ as a function of $M_D$ and $M_R$, by requiring disentanglement:
\begin{equation}\label{detB}
\sqrt{\mathbbm{1} - B^T B^*}\, M_D \sqrt{\mathbbm{1} - B\, B^\dagger} -
B^T M_D^T B^\dagger - \sqrt{\mathbbm{1} - B^T B^*} M_R B^\dagger = 0.
\end{equation}
By expanding $B$ in $m_D/m_R$, Eq.~(\ref{detB}) allows for a
recursive solution. 
One can show~\cite{disent} that 
$B = B_1 + B_3 + B_5 + \cdots$, where $B_n$ is of order $(m_D/m_R)^n$.
The first term in $B$ is easily
obtained from Eq.~(\ref{detB}) as 
\begin{equation}\label{B1}
B_1 = \left( M_R^{-1} M_D \right)^\dagger,
\end{equation}
and up to second order~\cite{schechter82}
\begin{equation}
W \simeq \left( \begin{array}{cc}
1 - \frac{1}{2}\, B_1 B_1^\dagger & B_1 \\ 
-B_1^\dagger & 1 - \frac{1}{2}\, B_1^\dagger B_1
\end{array} \right).
\end{equation}
Then one finds the leading term of the light-neutrino mass
matrix~\cite{seesaw}:  
\begin{equation}\label{seesaw-formula}
\mathcal{M}_\nu = -M_D^T M_R^{-1} M_D.
\end{equation}
This is the famous seesaw formula. In leading order, the 
mass matrix of heavy neutrinos is given by
\begin{equation} 
\mathcal{M}_\nu^\mathrm{heavy} = M_R.
\end{equation}
An interesting feature is that 
corrections to $\mnu$ and $\mnu^\mathrm{heavy}$ suppressed by $(m_D/m_R)^2$,
i.e., by the square of the small ratio $m_D/m_R$~\cite{disent}. This is a
consequence of the zero in the upper left corner of $\mathcal{M}_{D+M}$---see
Eq.~(\ref{D+M}). 

We note that the whole procedure sketched here goes through with an arbitrary
number of right-handed neutrinos. We could choose for instance two or more
than three $\nu_R$. However, if we chose only one right-handed neutrino, then
two neutrino masses are zero, which is in disagreement with the two non-zero
mass-squared differences; in this case an additional contribution to the
neutrino masses has to be supplied from elsewhere.

Let us for the moment assume that the charged-lepton mass matrix is
non-diagonal and it gets diagonalized by 
$(U_R^\ell)^\dagger M_\ell U_L^\ell = \hat m_\ell$.
In that case the lepton mixing matrix is given by 
$U_M = (U_L^\ell)^\dagger V$ and there are \emph{three sources for mixing:}
$M_\ell$, $M_D$ and $M_R$. Thus the 
seesaw mechanism is a rich playground for model building.

To conclude the seesaw part, we want make a scale consideration. Choose as a
typical neutrino mass 
$m_\nu \sim \sqrt{\Delta m^2_\mathrm{atm}} \sim 0.05$ eV and 
$m_D \sim m_{\mu,\tau}$. Then, $m_R \sim 10^8 \div 10^{11}$ GeV. 
This is fairly close to the GUT scale. 
Could the seesaw scale $m_R$ be identical with a GUT scale of typically
$M_\mathrm{GUT} \simeq 2 \times 10^{16}$ GeV? 
This is not possible without some amount of finetuning because 
$m_D \lesssim v$, where the VEV $v \simeq 174$ GeV represents the electroweak
scale. Then, according to the seesaw mechanism 
$m_\nu \sim v^2/M_\mathrm{GUT} \sim 1.5 \times 10^{-3}$ eV, 
which is too small. On the other hand, in the minimal SUSY extension of the SM 
gauge coupling unification happens at $M_\mathrm{GUT}$ and the question is if
in a certain GUT model an intermediate scale like the seesaw scale is allowed.
It has been shown that the so-called minimal SUSY $SO(10)$ GUT is ruled out
for this reason~\cite{MSGUT}. This problem is an interesting research topic. 

\subsection{Additional scalar multiplets}

The leptonic SM multiplets are characterized in the beginning of
Section~\ref{right-handed}. 
With these multiplets one can form 
leptonic bilinears with the following quantum numbers~\cite{KK}:
$$
\begin{array}{cr@{\,=\,}lccl}
\bar D_L \otimes \ell_R & 
\underline{\frac{1}{2}} \otimes \underline{0} & \underline{\frac{1}{2}} &
Y = -1 & \:\phi & \mbox{doublet}\,\, Y = +1, \\
D_L \otimes D_L & 
\underline{\frac{1}{2}} \otimes \underline{\frac{1}{2}} & 
\underline{0} \oplus \underline{1} & Y = -2 &
\left\{ \begin{array}{c} \eta^+ \\ \!\!\Delta \end{array} \right. &
\!\!\!\begin{array}{l} \mbox{singlet} \,\,\,\, Y = +2, \\ 
\mbox{triplet} \,\,\,\, Y = +2, \end{array} \\
\ell_R \otimes \ell_R & \underline{0} \otimes \underline{0} &
\underline{0} & Y = -4 & \quad k^{++} & \mbox{singlet} \,\,\, Y = +4.
\end{array}
$$
This is the complete SM list. One can 
add $\nu_R \otimes \nu_R$ with trivial quantum numbers 
$\underline{0} \otimes \underline{0} = \underline{0}, \: Y = 0$, which
tells us that this bilinear can couple to a real or complex scalar singlet.
We will not consider this possibility in these lecture notes.

In the list above, apart from the trivial case with weak isospin
$1/2$, there are three interesting new scalar multiplets, which enable 
massive Majorana neutrinos. 
Three such models, corresponding to these multiplets,  
will be discussed in the following.

\paragraph{The Zee model:} 
The essential ingredient of this model is the charged scalar
$\eta^+$~\cite{zee1,zee2}. In addition to the SM multiplets,
it needs a second Higgs doublet. The relevant parts of the Lagrangian 
are
\begin{equation}\label{Lzee}
\mathcal{L} = \cdots 
+ f_{\alpha\beta} D_{\alpha L}^T C^{-1} i\tau_2 D_{\beta L}\, \eta^+ -
\mu\, \phi_1^\dagger \tilde \phi_2 \eta^+ + \mbox{H.c.},
\end{equation}
where $f_{\alpha\beta}$ is the Yukawa coupling matrix of $\eta^+$.
Without loss of generality we can assume that the charged-lepton mass
matrix is diagonal and, therefore, $\alpha$ and $\beta$ indicate the
flavour, i.e., $\alpha,\, \beta = e,\, \mu,\, \tau$. The $\eta^+$
Yukawa interaction has a ``Majorana'' structure and is
Lorentz-invariant just as the Majorana mass term discussed in
Section~\ref{majorana}. The Pauli matrix $\tau_2$ acts on the $SU(2)$
doublets $D_{\beta L}$, i.e., it has $SU(2)$
indices. Equation~(\ref{su2prop}) expresses the invariance of the $\eta^+$
Yukawa interaction under $SU(2)$.

In Section~\ref{majorana} we have derived that a Majorana mass matrix
is symmetric---see Eq.~(\ref{symmetric}). We can apply an analogous
reasoning here, however, due to the antisymmetry of $\tau_2$, we have
\begin{equation}
D_{\alpha L}^T C^{-1} i\tau_2 D_{\beta L} = -
D_{\beta L}^T C^{-1} i\tau_2 D_{\alpha L}
\quad \mbox{and} \quad 
f_{\alpha\beta} = -f_{\beta\alpha}.
\end{equation}

Why does the Zee model need two Higgs doublets? 
The Zee model aims at radiative neutrino mass generation. Since there
is no $\nu_R$, the neutrinos have to be of the Majorana type, with
violation of the total lepton number $L$. Therefore, a necessary
condition for non-zero neutrino masses in the Zee model is
$L$-violation. Considering the $\eta^+$
Yukawa interaction in Eq.~(\ref{Lzee}) and the Yukwawa interactions 
of the Higgs doublets, leads us to the following lepton number
assignment: 
$$
\begin{array}{c|cccc}
  & D_L & \ell_R & \phi_{1,2} & \eta^+ \\ \hline
L & 1   & 1      & 0          & -2
\end{array}
$$
Clearly, such a lepton number is explicitly broken by the $\mu$-term in
Eq.~(\ref{Lzee}), and this is the only term in the total Lagrangian
which breaks $L$. Suppose we have only one Higgs doublet, i.e., 
$\phi_1 = \phi_2 \equiv \phi$. Then it follows that 
$\phi_1^\dagger \tilde \phi_2 = \phi^\dagger \tilde \phi \equiv 0$ 
and the $\mu$-term is absent and
without it the neutrinos remain massless.

\begin{figure}
\begin{center}
\scalebox{0.5}{\includegraphics*[5cm,15cm][17cm,25cm]{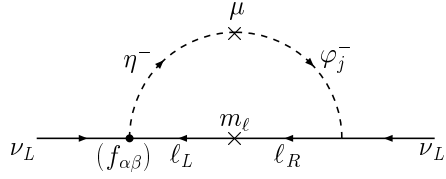}}
\end{center}

\vspace{-17mm}

\caption{The 1-loop Feynman diagram for the generation of neutrino masses
  in the Zee model. \label{zeefig}}
\end{figure}

A restricted version of Zee model has been proposed by
Wolfenstein~\cite{wolfenstein-zee}. In that version 
only $\phi_1$ couples to leptons. This is guaranteed if we introduce
the symmetry 
\begin{equation}
D_L \to iD_L, \quad \ell_R \to i\ell_R, \quad \phi_1 \to \phi_1, \quad 
\phi_2 \to -\phi_2, \quad \eta^+ \to -\eta^+.
\end{equation}
Note that in this case the Yukawa coupling matrix of $\phi_1$ is 
proportional to the diagonal mass matrix of the charged leptons. 
The 1-loop Feynman diagram which generates neutrino masses is depicted
in Fig.~\ref{zeefig}.
The vertex at the right corner of the diagram has a charged-lepton
mass from the Yukawa coupling, another one comes from the mass
insertion on the fermion line. 
The diagram describes a transition $\nu_L \to \left( \nu_L \right)^c$,
in accord with a Majorana mass term---see Eq.~(\ref{majterm}).
Taking into account that $\mnu$ is
symmetric, one finds~\cite{zee1}
\begin{equation}\label{zee-rest}
\mathcal{M}_\nu \propto \left( (m_\alpha^2 - m_\beta^2)
  f_{\alpha\beta} \right) 
\quad \mbox{or} \quad 
\mnu = \left( \begin{array}{ccc}
0 & a & b \\ a & 0 & c \\ b & c & 0 \end{array} \right).
\end{equation}
Thus the restricted version of the Zee model generates the 
most general Majorana mass matrix with zeros on the diagonal. 
After removal of unphysical phases, one obtains a real 3-parameter
mass matrix. 
However, the mass matrix~(\ref{zee-rest}) is not
viable~\cite{jarlskog}, because it predicts that
solar mixing is, for all practical purposes, maximal. 

Thus the restricted version is ruled out. However, if 
both Higgs doublets are allowed to couple in the lepton sector,
i.e., there are two different coupling matrices at the right corner of
the diagram in Fig.~\ref{zeefig}, then $\mnu$ has in general non-zero
entries in the diagonal, non-maximal solar mixing is allowed and 
there is no contradiction with experimental results~\cite{balaji}. 
On the other hand, though neutrino masses are
suppressed because they are generated radiatively, further suppression
is required to get neutrino masses of order 1 eV. If we perform such a
suppression by small Yukawa couplings of the $\eta^+$, a rough estimate is 
$|f_{\alpha\beta}| \lesssim 10^{-4}$.

\paragraph{The Zee--Babu model:}
In this model~\cite{zee2,babu}, the SM multiplets are enriched by the
scalar singlets $\eta^+$ and $k^{++}$. 
The relevant parts of the Lagrangian are given by
\begin{equation}\label{Lbabu}
\mathcal{L} = \cdots 
+ f_{\alpha\beta} D_{\alpha L}^T C^{-1} i\tau_2 D_{\beta L}\, \eta^+ +
h_{\alpha\beta} \ell_{\alpha R}^T C^{-1} \ell_{\beta R}\, k^{++} - 
\tilde\mu\, \eta^-\eta^- k^{++} + \mbox{H.c.}
\end{equation}
Now we use arguments similar to those for the Zee model. 
The Yukawa coupling matrix of $k^{++}$ has the property
\begin{equation}
h_{\alpha\beta} = h_{\beta\alpha}.
\end{equation}
If we assign lepton numbers $L$, then $L(k^{++}) = -2$, 
but $L$ is explicitly broken by the $\tilde\mu$-term. Thus,
we obtain radiative neutrino masses, with neutrinos of the
Majorana type.

\begin{figure}
\begin{center}
\scalebox{0.5}{\includegraphics*[5cm,17cm][17cm,24cm]{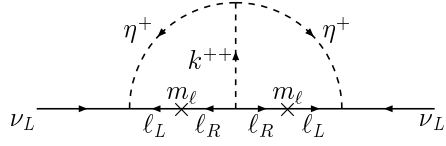}}
\end{center}

\vspace{-17mm}

\caption{The 2-loop Feynman diagram for the generation of neutrino
  masses in the Zee--Babu model. \label{babufig}}
\end{figure}

The Zee--Babu model has only one Higgs doublet. According to the
discussion of the Zee model, at 1-loop the neutrino masses are still
zero, but they appear at the 2-loop level. The relevant Feynman
diagram is depicted in Fig.~\ref{babufig}. Examination of that diagram
shows that
\begin{equation}
\mathcal{M}_\nu \propto \tilde f \hat m_\ell {\tilde h}^* \hat m_\ell
\tilde f,
\end{equation}
with
$\tilde f = (f_{\alpha\beta})$, $\tilde h = (h_{\alpha\beta})$ and 
$\hat m_\ell = \mbox{diag}\,(m_e,m_\mu,m_\tau)$.

The properties of the model are the following. Since 
$\tilde f$ is antisymmetric, the lightest neutrino mass is zero. 
Assuming neutrino mass hierarchy requires the
fine-tuning~\cite{macesanu} 
$|h_{\mu\mu}| : |h_{\mu\tau}| : |h_{\tau\tau}| \simeq 
1 : (m_\mu/m_\tau) : (m_\mu/m_\tau)^2$. 
With scalar masses in the TeV range, small neutrino masses require
$|f_{\alpha\beta}|$, $|h_{\alpha\beta}| \lesssim 0.1$. Thus, with
2-loop suppression, neutrino masses turn out to be naturally small.
As a bonus, 
rare decays like $\tau \to 3\mu$ and $\mu \to e \gamma$ are within
reach of forthcoming experiments.

We want to finish our discussion of the two models for 
radiative neutrino mass generation with a remark. 
Via the loop diagrams the hierarchy of the charged-lepton
masses is transferred into the neutrino mass matrix, which is, therefore,
naturally of hierarchical structure. Thus, finetuning of Yukawa
couplings seems to be unavoidable in order to reproduce large lepton
mixing. 

\paragraph{The triplet model:}
That model is obtained by enlarging the SM by a scalar triplet $\Delta$.
This triplet is conveniently written as a $2 \times 2$ matrix with
$SU(2)$ indices, parameterized by
\begin{equation}\label{delta}
\Delta = \sum_{j=1}^3 \delta_j \tau_j,
\end{equation}
where the $\tau_j$ are the Pauli matrices. 
The relevant terms of the Lagrangian are given by~\cite{gelmini}
\begin{equation}\label{Ltriplet}
\mathcal{L} = \cdots 
+ \frac{1}{2} g_{\alpha\beta} D_{L\alpha}^T C^{-1} i\tau_2 \Delta
D_{L\beta} + \mbox{H.c.} - M^2 \mbox{Tr}\, \Delta^\dagger \Delta - 
\left( \mu_\Delta\, \phi^\dagger \Delta \tilde \phi + \mbox{H.c.} \right)
- \cdots
\end{equation}
The first dots indicate the gauge part, the second ones the
missing terms of the scalar potential which are not relevant for our
discussion. 
Observing that the $\tau_2 \tau_j$ are symmetric for $j=1,2,3$, we find
\begin{equation}
g_{\alpha\beta} = g_{\beta\alpha}.
\end{equation}

The triplet coupling is $SU(2)$-invariant for the following reason. 
With $U \in SU(2)$, the transformation properties are 
\begin{equation}\label{transformation}
D_L \to U D_L, \quad \Delta \to U \Delta U^\dagger.
\end{equation}
Using the parameterization~(\ref{delta}), the latter equation
demonstrates that $\vec \delta$ transforms according to the adjoint
representation of $SU(2)$, i.e., as an $SU(2)$ triplet. 
Invariance under $SU(2)$ of the triplet term is again proved by using 
Eq.~(\ref{su2prop}). 

Let us now determine the charge-eigenfields of the triplet. First we
note that its hypercharge is $Y_\Delta = 2$. 
In general, the electric charge in a multiplet of the SM gauge group
is given by
$\mathbbm{Q} = T_3 + \frac{1}{2}\,Y$, where $T_3$ is the third
$SU(2)$ generator, i.e., the third component of weak
isospin. According to Eq.~(\ref{transformation}), we thus obtain 
\begin{equation}
\mathbbm{Q} \Delta = \frac{1}{2}\,[\tau_3, \Delta] + \Delta
\end{equation}
or
\begin{equation}
\left( \begin{array}{cc} \delta_3 & \delta_1 - i\delta_2 \\
\delta_1 + i\delta_2 & -\delta_3 
\end{array} \right) \stackrel{\mathbbm{Q}}{\to} 
\left( \begin{array}{cc} \delta_3 & 2 \left(\delta_1 - i\delta_2 \right) \\
0 \left( \delta_1 + i\delta_2 \right) & -\delta_3 
\end{array} \right).
\end{equation}
Therefore, 
\begin{equation}\label{triplet}
\Delta = \left( \begin{array}{cc} H^+ & \sqrt{2} H^{++} \\
\sqrt{2} H^0 & -H^+ \end{array} \right),
\end{equation}
with
$H^0 = \frac{1}{\sqrt{2}} \left( \delta_1 + i\delta_2 \right)$, 
$H^+ = \delta_3$, 
$H^{++} = \frac{1}{\sqrt{2}} \left( \delta_1 - i\delta_2 \right)$.

Assigning lepton numbers as in the previous models, 
we find $L(\Delta) = -2$, and $L$ is explicitly broken by the
$\mu_\Delta$-term. However, according to Eq.~(\ref{triplet}),
the triplet has a neutral component which can have a VEV
\begin{equation}
\langle H^0 \rangle_0 = \frac{1}{\sqrt{2}} v_T.
\end{equation}
Thus in the triplet model there is a  
tree-level neutrino mass matrix~\cite{gelmini}
\begin{equation}
\mathcal{M}_\nu = v_T (g_{\alpha\beta}).
\end{equation}
Since the triplet VEV disturbs the famous tree level relation
$M_W/M_Z = \cos \theta_W$ for $W$ mass, $Z$ mass and Weinberg angle
$\theta_W$, there is an upper bound $|v_T|/v \lesssim 0.03$ from the
LEP data~\cite{erler}. 
However, in order to have small enough neutrino masses from a 
small VEV, $v_T$ must be much smaller, namely
$v_T \sim 0.1 \div 1$ eV.

How to get a small $v_T$? Mechanisms in analogy to the seesaw
mechanism, where the order of the neutrino masses is given by $m_D^2/m_R$
with $m_D \ll m_R$, have been proposed such that the triplet VEV is
obtained by an analogous order of magnitude relation. Such mechanisms
are usually called type~II seesaw~\cite{II} (see also~\cite{pal}).
In that context the original seesaw mechanism is called type~I.
Here we follow Ref.~\cite{ma}. We assume
\begin{equation}
M, \: |\mu_\Delta| \gg v,
\end{equation}
where $v \simeq 174$ GeV is the VEV of the Higgs doublet.
Replacing in the scalar potential of Eq.~(\ref{Ltriplet})
the fields by the VEVs, we obtain
\begin{equation}
\langle V \rangle_0 = M^2 v_T^* v_T + 
v^2 \mu_\Delta v_T + v^2 \mu^* v_T^* + \cdots
\end{equation}
The dots contain terms of order $v^4$ and $v_T^2 v^2$. 
The stability condition with respect to $v_T$ requires 
\begin{equation}
\frac{\partial}{\partial v_T^*} \langle V \rangle_0 = 
M^2 v_T + v^2 \mu^* + \mathcal{O}(v_T v^2) = 0,
\end{equation}
which results in
\begin{equation}
|v_T| \simeq |\mu| v^2/M^2.
\end{equation}
Indeed, $v_T$ is of the form $v^2/m_S$, where $m_S$ is a large scale
in the scalar sector.
We want to emphasize that the mechanism for generating a small non-zero
$v_T$ has a peculiar feature. It requires $M^2 > 0$, thus without the
VEV of the Higgs doublet the triplet VEV would be zero. On the other
hand, the VEV of $\phi$ is non-zero because in the scalar potential
the term $\phi^\dagger \phi$ has a \emph{negative} coefficient.

In summary, in the triplet model, generation of small neutrino masses
requires a small triplet VEV $v_T$, which in turn requires a new 
new heavy scale in scalar sector, in analogy
to the seesaw mechanism of type~I. 
Seesaw mechanisms of type~I+II are naturally obtained in $SO(10)$
GUTs, see for instance~\cite{pal}.
If both types are together, the mass matrix~(\ref{D+M}) looks like 
\begin{equation}
\setlength{\arraycolsep}{2pt}
\mathcal{M}_{D+M} = \left( \begin{array}{cc}
M_L & M_D^T \\ M_D & M_R     \end{array} \right),
\end{equation}
with the seesaw formula
\begin{equation}
\mathcal{M}_\nu = M_L -M_D^T M_R^{-1} M_D.
\end{equation}
Note that now all terms in $B = B_1 + B_2 + B_3 + \cdots$ in an
expansion in $1/m_R$ appear---see for instance~\cite{disent}, for a
thorough discussion. The matrix $M_L$ need not necessarily be present
at tree level, as it the case in the triplet model, but can also
appear via loop corrections in some models---see for instance~\cite{1loop}.

\section{The $\mu$--$\tau$-symmetric neutrino mass matrix}
\label{mu-tau-section}

We depart from the Majorana mass term~(\ref{mnu}) and the 
basis where charged-lepton mass matrix is diagonal.
We consider the mass matrix
\begin{equation}\label{mu-tau}
\mathcal{M}_\nu = \left( \begin{array}{ccc} x & y & y \\ y & z & w \\
y & w & z \end{array} \right)
\quad \mbox{with} \quad
x, \, y, \, z, \, w \in \mathbbm{C},
\end{equation}
which is symmetric under $\mu$--$\tau$ interchange---for early
references on this mass matrix see~\cite{early,GL01}.

The $\mu$--$\tau$ interchange symmetry in $\mnu$ can be defined
as~\cite{maximal} 
\begin{equation}\label{S}
S\mathcal{M}_\nu S = \mathcal{M}_\nu 
\quad \mbox{with} \quad
S =
\left( \begin{array}{ccc} 1 & 0 & 0 \\ 0 & 0 & 1 \\ 0 & 1 & 0
\end{array} \right).
\end{equation}

Let us now discuss the phenomenology of the mass
matrix~(\ref{mu-tau}). 
We immediately guess one eigenvector: 
\begin{equation}
\left( \begin{array}{ccc} x & y & y \\ y & z & w \\
y & w & z \end{array} \right) 
\left( \begin{array}{r} 0 \\ 1 \\ -1 \end{array} \right) =
(z - w) 
\left( \begin{array}{r} 0 \\ 1 \\ -1 \end{array} \right).
\end{equation}
Since this vector is real, it can be identified with one of the
columns in the diagonalization matrix $V$, defined in Eq.~(\ref{V}).
The first component of this eigenvector is zero, thus we identify it
with the third column of $V$ because $|V_{e3}| = s_{13}$ (see
Eqs.~(\ref{V1}) and (\ref{U})) is
compatible with zero. Therefore, we also obtain 
$m_3 = | z - w |$. Inspecting $U$ of Eq.~(\ref{U}) we find with the
eigenvector above that 
\begin{equation}\label{angles}
\theta_{23} = 45^\circ, \quad \theta_{13} = 0^\circ,
\end{equation}
and the parameter measured in atmospheric and long-baseline
experiments is maximal, i.e.,
\begin{equation}
\sin^2 2\theta_\mathrm{atm} = 
4\, |U_{\mu 3}|^2 \left( 1 - |U_{\mu 3}|^2 \right) = 1.
\end{equation}
Because of Eq.~(\ref{angles}), the matrix $V$ is given by 
\begin{equation}\label{V2}
V = 
\mathrm{diag}\, (e^{i\varphi_e}, e^{i\varphi_\mu}, -e^{i\varphi_\mu}) 
\left( \begin{array}{ccc}
\scriptstyle \cos \theta & \scriptstyle \sin \theta & \scriptstyle 0 \\
-\frac{\sin \theta}{\sqrt{2}} & \frac{\cos \theta}{\sqrt{2}} 
& \frac{1}{\sqrt{2}} \\
-\frac{\sin \theta}{\sqrt{2}} & \frac{\cos \theta}{\sqrt{2}} 
& -\frac{1}{\sqrt{2}}
\end{array} \right)
\mathrm{diag}\, (e^{i\rho}, e^{i\sigma}, 1).
\end{equation}
That the diagonal phase matrix to the left of $U$ has this specific
form, can be proved from the $\mu$--$\tau$ interchange symmetry. Note
we have slightly changed the convention of $U$ compared to
Eq.~(\ref{U}): the third line has been multiplied by $-1$.

Equation~(\ref{angles}) represents the predictions of the mass
matrix~(\ref{mu-tau}) and implies that it 
is compatible with all data. 
If one can generate such a mass matrix in a
model by means of symmetries, on would have an explanation for
large atmospheric mixing and small $\theta_{13}$. Obviously,
the $\mu$--$\tau$-symmetric mass matrix is more specific than that and
will be tested by future experimental efforts.

Since $s_{13} = 0$, the CP phase $\delta$ is meaningless. The mass
matrix~(\ref{mu-tau}) has no predictions for the masses; they are free
and all types of mass spectra are admitted. 
The $\mu$--$\tau$-symmetric mass matrix is thus an example of what was
mentioned in the introduction: there might be mass-independent
predictions for mixing angles and the mass problem is deferred to a
more fundamental theory.

In the neutrino sector, we have nine observables: three neutrino
masses, three mixing angles and three CP phases. If we remove the
unphysical phases from the mass matrix~(\ref{mu-tau}), for instance by
making the first row and first column real by a phase transformation,
we are left with six parameters. That mass matrix gives two
predictions---see Eq.~(\ref{angles})---and the phase $\delta$ drops
out from $U$. Thus we are left with six observables, the masses, the
solar mixing angle and the Majorana phases, which are not predicted
by Eq.~(\ref{mu-tau}); this is in agreement with the six physical parameters
in $\mnu$ counted above.

One can ask the question if a small perturbation of the
$\mu$--$\tau$-symmetric mass matrix destroys the
predictions~(\ref{angles}). It turns out that they stable for
$m_{1,2,3} \lesssim \sqrt{\Delta m_\mathrm{atm}^2}$~\cite{sawanaka}. The
predictions are unstable for a degenerate neutrino mass spectrum where
$m_{1,2,3} \gg \sqrt{\Delta m_\mathrm{atm}^2}$.

The matrix $U$ in Eq.~(\ref{V2}) may be further specified by
fixing the solar mixing angle by~\cite{HPS}
\begin{equation}
\sin^2 \theta = \frac{1}{3}, 
\end{equation}
which leads to the so-called tri-bimaximal mixing matrix
\begin{equation}
\left( \begin{array}{ccc}
\scriptstyle \cos \theta & \scriptstyle \sin \theta & \scriptstyle 0 \\
-\frac{\sin \theta}{\sqrt{2}} & \frac{\cos \theta}{\sqrt{2}} 
& \frac{1}{\sqrt{2}} \\
-\frac{\sin \theta}{\sqrt{2}} & \frac{\cos \theta}{\sqrt{2}} 
& -\frac{1}{\sqrt{2}}
\end{array} \right)
=
\left( \begin{array}{ccc}
\frac{2}{\sqrt{6}} & \frac{1}{\sqrt{3}} & 0 \\
-\frac{1}{\sqrt{6}} & \frac{1}{\sqrt{3}}
& \frac{1}{\sqrt{2}} \\
-\frac{1}{\sqrt{6}} & \frac{1}{\sqrt{3}}
& -\frac{1}{\sqrt{2}}
\end{array} \right).
\end{equation}
Here the solar mixing angle is $\theta \simeq 35.3^\circ$, which is in
good agreement with the experimentally allowed range. 
Models which explain tri-bimaximal mixing exist, but they are quite
involved~\cite{theory-reviews}.

\section{A model for the $\mu$--$\tau$-symmetric neutrino mass matrix}
\label{model}

\paragraph{Multiplets and symmetries:} 
The model to be discussed here was published in~\cite{GL01}. Its 
multiplets are those of the SM, however, it needs three Higgs doublets
$\phi_j$ ($j=1,2,3$), and the fermion sector contains in addition
three right-handed neutrino singlets, in order to generate light
neutrino masses by the seesaw mechanism. 
The symmetries and symmetry breakings of the model are the following:
\begin{itemize}
\item[$\rhd$]
Three groups $U(1)_{L_\alpha}$ ($\alpha = e,\mu,\tau$), associated
with the family lepton numbers $L_\alpha$, which are 
\emph{softly} broken by the $\nu_R$ mass term (note that a fermion
mass term has dimension three);
\item[$\rhd$]
$\mathbbm{Z}_2^{(\mathrm{tr})}$:
$D_{\mu L} \leftrightarrow D_{\tau L}$, 
$\mu_R \leftrightarrow \tau_R$, 
$\nu_{\mu R} \leftrightarrow \nu_{\tau R}$, 
$\phi_3 \to - \phi_3$,
which is spontaneously broken by the VEV of $\phi_3$;
\item[$\rhd$]
$\mathbbm{Z}_2^{(\mathrm{aux})}: \;
\nu_{eR},\: \nu_{\mu R},\: 
\nu_{\tau R},\: \phi_1,\:
e_R\, \: \mbox{change sign}$, which is 
spontaneously broken the VEV of $\phi_1$.
\end{itemize}
These symmetries determine the Yukawa Lagrangian:
\begin{eqnarray}
\mathcal{L}_\mathrm{Y} & = & 
- y_1 \bar D_{eL} \nu_{eR} \tilde\phi_1  
- y_2 \left( \bar D_{\mu L} \nu_{\mu R} + \bar D_{\tau L} \nu_{\tau R} \right)
\tilde\phi_1 
\nonumber \\ && \label{Lyukawa}
- y_3 \bar D_{eL} e_R \phi_1
- y_4 \left( \bar D_{\mu L} \mu_R + \bar D_{\tau L} \tau_R \right) \phi_2
- y_5 \left( \bar D_{\mu L} \mu_R - \bar D_{\tau L} \tau_R \right) \phi_3
+ \mbox{H.c.}
\end{eqnarray}
By virtue of the family lepton numbers $L_\alpha$, all Yukawa coupling
matrices are diagonal.

\paragraph{The neutrino mass matrix:}
Consider the mass term 
\begin{equation}\label{MR}
\mathcal{L}_M = \frac{1}{2}\, \nu_R^T C^{-1} \! M_R^* \nu_R +
\mbox{H.c.}
\end{equation}
of the right-handed neutrinos. According to the way we have stipulated
our symmetries, $\mathcal{L}_M$ is invariant under 
$\mathbbm{Z}_2^{(\mathrm{tr})}$ but not under the $U(1)_{L_\alpha}$.
Thus we find 
\begin{equation}
M_D = \mbox{diag}\, (a,b,b) \quad \mbox{and} \quad
M_R = \left( \begin{array}{ccc} m & n & n \\ n & p & q \\ n & q & p
\end{array} \right).
\end{equation}
Using the matrix $S$ of Eq.~(\ref{S}), 
the $\mathbbm{Z}_2^{(\mathrm{tr})}$ invariance translates into 
\begin{equation}
S M_D S = M_D, \quad S M_R S = M_R.
\end{equation}
But then we also have $S \mnu S = \mnu$, because $\mnu$ is obtained
from $M_D$ and $M_R$ by 
the seesaw formula~(\ref{seesaw-formula}). Thus the model
discussed here provides us with the $\mu$--$\tau$-symmetric neutrino
mass matrix discussed in the previous section and has, therefore,
the predictions of Eq.~(\ref{angles}).

Two remarks are at order. As mentioned in the previous section, the
masses are free in this model, therefore, we have no explanation for
$\Delta m^2_\odot/\Delta m^2_\mathrm{atm} \sim 0.03$. However, this
ratio is not very small and is a function of the elements in $M_D$
and $M_R$. Due to the seesaw mechanism, the entries in $\mnu$ are
sums over a product of four such elements times the common factor
$1/\det M_R$. Thus the ratio of mass-squared differences can easily be
reproduced if the elements in $M_{D,R}$ differ by 
no more than factors of $2 \div 3$. 
The second remark refers to radiative corrections. 
The $\mu$--$\tau$-invariance of $\mnu$ is expected to hold at the 
seesaw scale. Evolution of $\mnu$ according to the renormalization
group from the high scale down to the low scale introduces deviations
from the predictions~(\ref{angles}); however, one can check that this
effect changes $\mnu$ very little~\cite{RGE}. 
Only for a degenerate spectrum with
a common mass $m_0 \simeq 0.2$ eV, $s_{13}$ is at most 0.1 at the
electroweak scale, whereas $\sin^2 2\theta_{23}$ remains very close to
one. However, the correction to $s_{13}$ goes roughly with $m_0^2$ and
becomes quickly small for smaller $m_0$.

\paragraph{The model and the group $O(2)$:}
The symmetries described before generate a non-abelian symmetry group
because $U(1)_{L_\mu} \times U(1)_{L_\tau}$ and 
$\mathbbm{Z}_2^{(\mathrm{tr})}$ 
do not commute, but they generate the group $O(2)$~\cite{su5}, 
as we will demonstrate now.

First we give an abstract characterization of $O(2)$, 
a symmetry group in the plane. 
It contains rotations $g(\omega)$ ($\omega \in \mathbbm{R}$) 
with the multiplication laws 
$g(\omega + 2\pi) = g(\omega)$, 
$g(\omega_1)\,g(\omega_2) = g(\omega_1 + \omega_2)$. 
Another element is the reflexion $s$ at the $x$-axis with $s^2 = e$, 
where $e$ is the unit element. 
The reflexion $s$ and the rotations
$g(\omega)$ generate $O(2)$, if we stipulate the 
relation
$s\,g(\omega) = g(-\omega)\,s$.

The irreducible representations of $O(2)$ are easily found. There are 
two 1-dimensional representations
\begin{equation}
\underline{1}:\; g(\omega) \to 1,\; s \to 1, \quad
\underline{1}':\; g(\omega) \to 1,\; s \to -1,
\end{equation}
and there is an 
infinite series of 2-dimensional irreducible representations 
characterized by $n \in \mathbbm{N}$:
\begin{equation}
\underline{2}^{(n)}: \quad
g(\omega) \to 
\left( \begin{array}{cc} e^{in\omega} & 0 \\ 0 & e^{-in\omega}
  \end{array} \right), \quad
s \to \left( \begin{array}{cc} 0 & 1 \\ 1 & 0 
  \end{array} \right).
\end{equation}

Now we make the identifications 
\begin{equation}
s \leftrightarrow \mathbbm{Z}_2^{(\mathrm{tr})}, \quad
g(\omega) \leftrightarrow e^{i\omega (L_\mu - L_\tau)}.
\end{equation}
Then 
$(D_{\mu L}, D_{\tau L})$, $( \mu_R, \tau_R )$, 
$(\nu_{\mu L}, \nu_{\tau L})$ are in $\underline{2}^{(1)}$, 
$\phi_3$ is in $\underline{1}'$, the remaining fields transform
trivially.
The full symmetry is group is then characterized by~\cite{su5}
\begin{equation}
U(1)_{L_e} \times U(1)_{(L_\mu + L_\tau)} \times 
O(2) \times \mathbbm{Z}_2^{(\mathrm{aux})}.
\end{equation}

\paragraph{The problem of $m_\mu \ll m_\tau$:} 
Actually, the $\mu$--$\tau$ interchange symmetry, as realized in the
symmetries of the model, would rather suggest that 
$| y_4 v_2| \sim | y_5 v_3|$ and, therefore, $m_\mu \sim
m_\tau$. However, we need 
\begin{equation}
m_\mu = | y_4 v_2 + y_5 v_3| \ll m_\tau = | y_4 v_2 - y_5 v_3|.
\end{equation}
A technical solution of this finetuning problem was given in~\cite{GL04}.

It consists in introducing a new symmetry
\begin{equation}\label{K}
K: \quad \mu_R \to -\mu_R, \quad \phi_2 \leftrightarrow \phi_3,
\end{equation}
without introducing new multiplets. This is an important point because
this makes the model simpler, since the number of free parameters is
reduced. Implementing $K$ in the Yukawa Lagrangian~(\ref{Lyukawa}), we
obtain
\begin{equation}
y_4 = -y_5 \quad \mbox{and} \quad
\frac{m_\mu}{m_\tau} = \left| \frac{v_2-v_3}{v_2+v_3} \right|.
\end{equation}

As we want to show now, 
implementation of $K$ in the Higgs potential leads to 
$v_2 = v_3$ and, therefore, $m_\mu = 0$. If we are able to break $K$
softly by terms of dimension two in the Higgs potential,  
we will have $m_\mu \neq 0$, and thus $m_\mu \ll m_\tau$ in a
technically natural way. 

Since we need soft $K$-breaking, we assume that the Higgs potential
consists of two terms,
\begin{equation}
V = V_\phi + V_\mathrm{soft},
\end{equation}
where $V_\phi$ is $K$-invariant. It is easy to check that
the soft $K$-breaking term of dimension two is unique:
\begin{equation}
V_\mathrm{soft} = \mu^2_\mathrm{soft}
\left( \phi_2^\dagger \phi_2 - \phi_3^\dagger \phi_3 \right).
\end{equation}
To proceed further we mention that \emph{all} $\mathbbm{Z}_2$
symmetries listed in the beginning of this section, are broken
spontaneously. This means that $V$ is invariant under separate sign
transformations of all $\phi_j$ ($j=1,2,3$). With this observation, we
easily find
\begin{eqnarray}
V_\phi
&=& -\mu^2_1 \phi_1^\dagger \phi_1
- \mu^2_2 \left( \phi_2^\dagger \phi_2 + \phi_3^\dagger \phi_3 \right)
\nonumber \\ & &
+ \lambda_1 \left( \phi_1^\dagger \phi_1 \right)^2
+ \lambda_2 \left[ \left( \phi_2^\dagger \phi_2 \right)^2
+ \left( \phi_3^\dagger \phi_3 \right)^2 \right]
\nonumber \\ & &
+ \lambda_3 \left( \phi_1^\dagger \phi_1 \right) 
\left( \phi_2^\dagger \phi_2 + \phi_3^\dagger \phi_3 \right)
+ \lambda_4 \left( \phi_2^\dagger \phi_2 \right)
\left( \phi_3^\dagger \phi_3 \right)
\nonumber \\ & &
+ \lambda_5 \left[
\left( \phi_1^\dagger \phi_2 \right) \left( \phi_2^\dagger \phi_1 \right)
+ \left( \phi_1^\dagger \phi_3 \right) \left( \phi_3^\dagger \phi_1 \right)
\right]
+ \lambda_6
\left( \phi_2^\dagger \phi_3 \right) \left( \phi_3^\dagger \phi_2 \right)
\nonumber \\ & &
+ \lambda_7 \left[ \left( \phi_2^\dagger \phi_3 \right)^2 + 
\left( \phi_3^\dagger \phi_2 \right)^2 \right]
\nonumber \\ & &
+ \lambda_8 \left[ \left( \phi_1^\dagger \phi_2 \right)^2
+ \left( \phi_1^\dagger \phi_3 \right)^2 \right]
+ \lambda_8^\ast \left[ \left( \phi_2^\dagger \phi_1 \right)^2
+ \left( \phi_3^\dagger \phi_1 \right)^2 \right].
\end{eqnarray}
All coupling constants are real except $\lambda_8$. 
We make the ansatz~\cite{GL04}
\begin{equation}\label{ansatz}
v_2 = u\, e^{i\alpha} \cos{\sigma}, \quad 
v_3 = u\, e^ {i\beta} \sin{\sigma}
\quad \mbox{with} \quad 
v_1 > 0, \quad u > 0.
\end{equation}
Defining $F_\phi =  \langle V_\phi \rangle_0$ and using
Eq.~(\ref{ansatz}), we obtain
\begin{eqnarray}
F_\phi &=&
- \mu^2_1 v_1^2 - \mu^2_2 u^2
+ \lambda_1 v_1^4 + \lambda_2 u^4
+ \left( \lambda_3 + \lambda_5 \right) v_1^2 u^2
\nonumber \\ & &
+ \left[ \tilde \lambda
- 4 \lambda_7 \sin^2{\left( \alpha - \beta \right)} \right] 
u^4 \cos^2{\sigma} \sin^2{\sigma}
\nonumber \\ & &
+ 2 \left|\lambda_8 \right| v_1^2 u^2 
\left[ \cos^2 \sigma\, \cos \left( \epsilon + 2\alpha \right)
+ \sin^2 \sigma\, \cos \left( \epsilon + 2 \beta \right) \right].
\end{eqnarray}
In $F_\phi$ we have defined
$\tilde \lambda = - 2 \lambda_2 + \lambda_4 + \lambda_6 + 2
\lambda_7$ and $\epsilon = \arg{\lambda_8}$. 
The following is easy to check:
\begin{center}
If 
$\tilde \lambda < 0$ and $\lambda_7 < 0$ $\Rightarrow$ 
the minimum of $F_\phi$ is at 
$\sigma = \pi/4$, $\alpha = \beta = (\pi - \epsilon)/2$.
\end{center}
Consequently, the minimum of $V_\phi$ has
\begin{equation}
v_2 = v_3 = \frac{1}{\sqrt{2}}\,
u\, e^{i \left. \left( \pi - \epsilon \right) \right/ 2}
\end{equation}
and $m_\mu = 0$. If we include soft $K$-breaking by adding
\begin{equation}
\langle V_\mathrm{soft} \rangle_0 = 
\mu^2_\mathrm{soft} u^2 \cos{2 \sigma}
\end{equation}
to $F_\phi$, we obtain
\begin{equation}
\cos 2\sigma = \frac{2 \mu^2_\mathrm{soft}}{\raisebox{-2pt}%
{${\tilde\lambda u^2}$}}
\quad \mbox{and} \quad
\frac{m_\mu}{m_\tau} = 
\frac{| \cos 2\sigma |}{1 + \sqrt{1 - \cos^2 2\sigma}}.
\end{equation}
If we choose $\mu^2_\mathrm{soft} \ll u^2$, 
we enforce $m_\mu/m_\tau \ll 1$, which is technically natural because 
in the limit $\mu^2_\mathrm{soft} \to 0$ the symmetry $K$ is conserved
and for $\mu^2_\mathrm{soft} \neq 0$ it is broken only softly.

\section{A general framework: The seesaw mechanism with soft
  lepton-number breaking} 
\label{multi}

The model we have introduced in the previous section has three Higgs
doublets, however, all Yukawa coupling matrices are diagonal. Thus the
family lepton numbers $L_\alpha$ ($\alpha = e,\mu,\tau$) are conserved
in all terms with dimension four in the Lagrangian, but they are
softly broken by the $\nu_R$ mass term---see Eq.~(\ref{MR}).
One could, therefore, envisage the 
general framework of a multi-Higgs-doublet SM, with $n_H$ Higgs doublets, 
the seesaw mechanism and soft $L_\alpha$ breaking. It has the
following features~\cite{GL02}:
\begin{itemize}
\item[$\ast$]
It is a renormalizable theory. 
In particular, lepton-flavour-changing amplitudes are finite.
\item[$\ast$] \
The family lepton numbers $L_\alpha$ (and the total lepton number) 
are softly broken by the mass term of the right-handed neutrino
singlets 
at the \emph{high} seesaw scale $m_R$.
\item[$\ast$]
All Yukawa coupling matrices are diagonal and thus also 
$M_\ell$ and $M_D$.
\item[$\ast$]
The mass matrix $M_R$---see Eq.~(\ref{MR})---is the only source
of lepton mixing. 
\end{itemize}
Apart from allowing for interesting models for neutrino masses and
mixing by imposing further symmetries, this framework is in itself
interesting. It has flavour-changing neutral interactions induced by
$M_R$, nevertheless, some flavour-changing processes do not decouple
for $m_R \to \infty$, provided $n_H > 1$. It is 
the scalar sector which is responsible for this non-decoupling.

\begin{figure}
\begin{center}
\scalebox{0.8}{\includegraphics*[0cm,20.5cm][17cm,24cm]{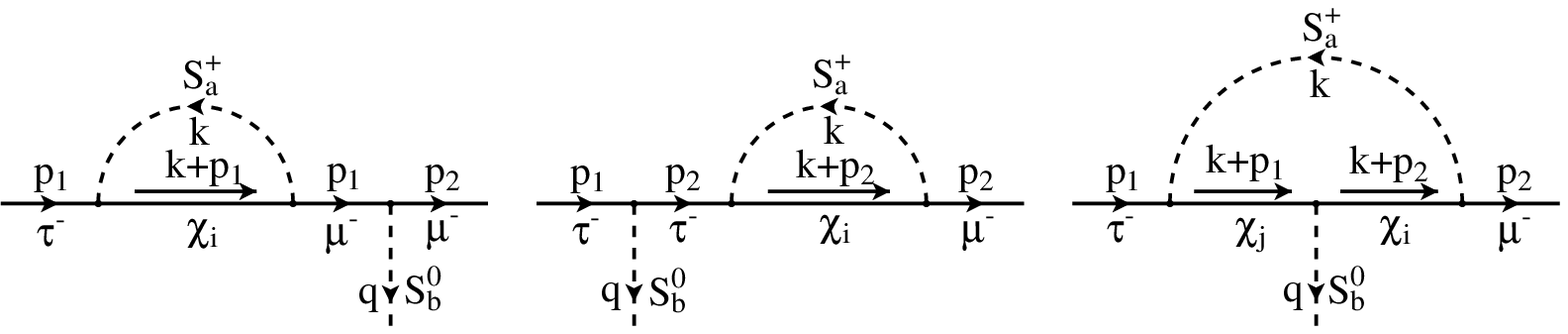}}
\end{center}
\caption{The diagrams for $\tau^- \to \mu^- \left( S_b^0 \right)^*$.
\label{unsuppressed}}
\end{figure}
Decays of charged leptons which are 
unsuppressed by $1/m_R$ are~\cite{GL02}
$\mu^- \to e^- e^+ e^-$, $\tau^- \to \mu^- e^+ e^-$,
$\tau^- \to \mu^- \mu^+ \mu^-$ and $\tau^- \to e^- e^+ e^-$. 
In Fig.~\ref{unsuppressed} we have depicted the Feynman diagrams 
for the vertex $\tau^- \to \mu^- \left( S_b^0 \right)^*$, where 
$\left( S_b^0 \right)^*$ is a virtual neutral scalar; this is the
vertex responsible for the amplitude 
$\mathcal{A}(\tau^- \to \mu^- \ell^+ \ell^-)$ ($\ell = e, \mu$)
which does not vanish in the limit $m_R \to \infty$. 
Also the flavour-changing scalar decays of the type 
$S_b^0 \to e^+ \mu^-$ are unsuppressed.
On the other hand, the decay amplitudes for
$\tau^- \to \mu^- \mu^- e^+$ and $\tau^- \to e^- e^- \mu^+$ 
stem from box diagrams and behave like $1/m_R^2$ for large $m_R$. The
amplitudes for $\mu \to e \gamma$ and $Z \to e^- \mu^+$ 
have the same behaviour.

While the processes whose amplitudes are suppressed by $1/m_R^2$ are
completely invisible for all practical purposes, 
the decay rate of $\mu^- \to e^- e^+ e^-$, 
although its amplitude is small because it contains a product of four
Yukawa couplings, could eventually be 
within experimental reach.

\vspace{5mm}

\noindent
\textbf{Acknowledgments:}
The author thanks the organizers for the invitation, the pleasant and
stimulating atmosphere and for choosing such a nice place for the school.

\end{document}